\documentclass[journal]{IEEEtran}
\usepackage[cmex10]{amsmath}
\usepackage{eqparbox}
\usepackage[usenames, dvipsnames]{color}
\usepackage{mathtools,amssymb,bm,mathabx}
\usepackage{amstext}
\usepackage{amssymb}
\usepackage{graphicx}
\usepackage{color}
\usepackage{booktabs}
\usepackage{longtable}
\usepackage{multicol}
\usepackage{amsfonts}
\usepackage{dsfont}
\usepackage{array}
\usepackage{algorithmicx}
\usepackage[ruled]{algorithm}
\usepackage{algpseudocode}
\usepackage{algpascal}
\usepackage{algc}
\usepackage{cases}
\usepackage{pgfplots}
\usepackage{accents}
\usepackage{caption}
\usepackage[footnotesize]{subfigure}
\usepackage{amsthm,xparse}
\DeclareCaptionLabelSeparator{periodspace}{.\quad}
\usepackage{float}
\usepackage{cite}
\usepackage [autostyle, english = american]{csquotes}
\usepackage{hyperref}
\theoremstyle{remark}

\newcommand\ASTART{\bigskip\noindent\begin{minipage}[b]{0.5\linewidth}}
	
	\newcommand\AENDSKIP{\end{minipage}\bigskip}
\newcommand\AEND{\end{minipage}}
\ifCLASSOPTIONcaptionsoff
\usepackage[nomarkers]{endfloat}
\let\MYoriglatexcaption\caption
\renewcommand{\caption}[2][\relax]{\MYoriglatexcaption[#2]{#2}}
\fi
\theoremstyle{plain}
\newtheorem{thm}{\textbf{Theorem}}

\newtheorem{prop}{\textbf{Proposition}}

\theoremstyle{definition}
\newtheorem{defn}{\textbf{Definition}}

\theoremstyle{remark}

\newcommand*{\rom}[1]{\expandafter\@slowromancap\romannumeral #1@}
\def\change{black}
\newcommand{\RN}[1]{%
\textup{\uppercase\expandafter{\romannumeral#1}}%
}

\usepackage{standalone}
\graphicspath{{./Figures/}} 
\begin{document}
%
\title{Improved Recovery of Analysis Sparse Vectors in Presence of Prior Information}
\author{Sajad~Daei, Farzan~Haddadi, Arash~Amini}%

\maketitle

\begin{abstract}
In this work, we consider the problem of recovering analysis-sparse signals from under-sampled measurements when some prior information about the support is available. We incorporate such information in the recovery stage by suitably tuning the weights in a weighted $\ell_1$ analysis optimization problem. Indeed, we try to set the weights such that the method succeeds with minimum number of measurements. For this purpose, we exploit the upper-bound on the statistical dimension of a certain cone to determine the weights. Our numerical simulations confirm that the introduced method with tuned weights outperforms the standard $\ell_1$ analysis technique.
\end{abstract}

\begin{IEEEkeywords}
$\ell_1$ analysis, prior information, conic integral geometry.
\end{IEEEkeywords}

%
\IEEEpeerreviewmaketitle

\section{Introduction}
 \IEEEPARstart{C}{ompressed} sensing (CS), initiated by \cite{candes2005decoding,donoho2006most}, has been the focus of many research works for more than a decade. Briefly, CS, in its general form, investigates the reconstruction of a sparse vector $\bm{x}\in\mathbb{R}^n$ from $m\ll n$  noisy linear measurements
 \begin{align}
 \bm{y}=\bm{A}\bm{x}+\bm{\epsilon}\in \mathbb{R}^m
 \end{align}
where $\bm{A}\in\mathbb{R}^{m\times n}$ is a known matrix and $\epsilon$ is an $\ell_2$ bounded noise term, i.e. $\|\epsilon\|_2\le \eta$ for some $\eta\ge0$. In many scenarios, $\bm{x}$ is sparse after the application of some analysis operator $\bm{\Omega}$. Specifically, we say $\bm{x}$ is s-analysis-sparse with support $\mathcal{S}$ in the analysis domain $\bm{\Omega}\in\mathbb{R}^{p\times n}$ if $\bm{\Omega x}$ is $s$-sparse with support $\mathcal{S}$. Then, the following optimization problem called $\ell_1$ analysis is often used (See \cite{candes2011compressed}, \cite{nam2013cosparse}, and \cite{kabanava2015analysis}) to recover $\bm{x}$:
\begin{align}\label{problem.l1analysis}
\mathsf{P}_{\eta}:~~\min_{\bm{z}\in \mathbb{R}^n}\|\bm{\Omega z}\|_1~~\mathrm{s.t.}~\|\bm{y}-\bm{A z}\|_2\le \eta
\end{align}
In many applications, there is some additional information in the analysis domain. For instance, consider the line spectral estimation where the signal of interest is sparse after applying the Discrete Fourier Transform. In some applications, one might a priori know the probability with which a set in the spectral domain contributes to the true line spectra. The extra information about the probability of contribution of certain subsets could be beneficial in the recovery of $\bm{x}$. For example, for channel estimation in communication systems or in remote sensing, the availability of previous estimates
builds a history that can specify the intersection probability of any given set with the true support. Also, natural images often tend to have larger values in lower frequencies after applying Fourier or wavelet transforms; therefore, subsets composed of low-frequencies have higher probabilities of appearing in the support. In these cases, we intend to exploit these additional information. This work analyses possible benefits of this extra information to reduce the required number of measurements of $\mathsf{P}_{0}$ for successful recovery and to improve the reconstruction error in $\mathsf{P}_{\eta}$ for robust and stable recovery. For this purpose, a common way is to use weighted $\ell_1$ analysis as follows:
\begin{align}\label{problem.weightedl1analysis}
&\mathsf{P}_{\eta}^{\bm{w}}:\nonumber\\
&~\min_{\bm{z}\in \mathbb{R}^n}\|\bm{\Omega z}\|_{1,\bm{w}}:=\sum_{i=1}^pw_i|\bm{\Omega z}
|_i~~\mathrm{s.t.}~\|\bm{y}-\bm{A z}\|_2\le \eta,
\end{align}
where $w_i$ represents the weight associated with the $i$'th element of the coefficient vector in the analysis domain. In this work, we assume that the available prior information is about the subsets $\{\mathcal{P}_{i}\}_{i=1}^L$ that partition $\{1,... ,p\}$. Thus, the elements of $\mathcal{P}_i$ are all assigned the same weight ($\omega_i$). Moreover, we define 
\begin{align}\label{eq.parameters}
\bm{w}=\sum_{i=1}^L{\omega}_i\bm{1}_{\mathcal{P}_i}~,\alpha_i=\frac{|\mathcal{P}_i\cap\mathcal{S}|}{|\mathcal{P}_i|},~ \rho_i=\frac{|\mathcal{P}_i|}{p},
\end{align}
where $|\cdot|$ denotes the cardinality of a set and $\bm{1}_{\mathcal{E}}$ is the indicator function of the set $\mathcal{E}$. The parameters $\alpha_i$ and $\rho_i$ are commonly called the accuracy and the normalized size of the subsets, respectively. Alternatively, $\{\mathcal{P}_i\}_{i=1}^L$ can be considered as $L$ analysis support estimators with different accuracies $\{\alpha_i\}_{i=1}^L$. Our goal is to find the weights that minimize the required number of measurements. To this end, we first find an upper-bound for the required number of measurements in Proposition \ref{prop.upperboundstatis}. Then, we minimize the upper-bound with respect to the weights. Since the bound is not tight (especially in redundant and coherent dictionaries), we can not claim optimality of the weights. However, with the obtained weights, we almost achieve the optimal phase transition curve of $\ell_1$ analysis problem in low-redundant analysis operators in numerical simulations. 
The paper is organized as follows: a brief overview of convex geometry is given in Section \ref{section.convexgeom}. We explain our main contribution in Section \ref{section.mainresult} followed by numerical experiments in Section \ref{section.simulation}. Indeed, the experiments confirm the theoretical results.

Throughout the paper, scalars are denoted by lowercase letters, vectors by lowercase boldface letters, and matrices by uppercase boldface letters. The $i$th element of a vector $\bm{x}$ is shown either by ${x}(i)$ or $x_i$. $(\cdot)^\dagger$ denotes the pseudo inverse operator. We reserve the calligraphic uppercase letters for sets (e.g. $\mathcal{S}$). The cardinality of a set $\mathcal{S}$ is denoted by $|\mathcal{S}|$. $\mathcal{C}^\circ$ represents the polar of a cone $\mathcal{C}$. Given a vector $\bm{x}\in\mathbb{R}^n$ and a set $\mathcal{C}\subseteq \mathbb{R}^n$, $\bm{x}\odot \mathcal{C}$ denotes the set $\mathcal{C}$ which is scaled by the elements of $\bm{x}$. In this work, $\bm{1}_{\mathcal{E}}$ denotes the indicator of the set $\mathcal{E}$. $(a)_+$ stands for $\max\{a,0\}$ for a scalar $a$. Null space and range of linear operators are denoted by $\mathrm{null}(\cdot)$, and $\mathrm{range}(\cdot)$, respectively. For a matrix $\bm{A}$, the operator norm is defined as $\|\bm{A}\|_{p\rightarrow q}=\underset{\|\bm{x}\|_p\le1}{\sup}\|\bm{Ax}\|_q$. We denote i.i.d standard Gaussian random vector by $\bm{g}$. Lastly, $\|\cdot\|_{\infty}$ returns the maximum absolute value of the elements of a vector or matrix.
\section{Convex Geometry}\label{section.convexgeom} 
In this section, basic concepts of conic integral geometry are reviewed.
\subsection{Descent Cones and Statistical dimension}
The descent cone of a proper convex function $f:\mathbb{R}^n\rightarrow \mathbb{R}\cup \{\pm\infty\}$ at point $\bm{x}\in \mathbb{R}^n$ is the set of directions from $\bm{x}$ that do not increase $f$:
\begin{align}\label{eq.descent cone}
\mathcal{D}(f,\bm{x})=\bigcup_{t\ge0}\{\bm{z}\in\mathbb{R}^n: f(\bm{x}+t\bm{z})\le f(\bm{x})\}\cdot
\end{align}
The descent cone of a convex function is a convex set. There is a famous duality \cite[Ch. 23]{rockafellar2015convex} between decent cone and subdifferential of a convex function  given by:
\begin{align}\label{eq.D(f,x)}
\mathcal{D}^{\circ}(f,\bm{x})=\mathrm{cone}(\partial f(\bm{x})):=\bigcup_{t\ge0}t.\partial f(\bm{x}).
\end{align}
\begin{defn}{Statistical Dimension}\cite{amelunxen2013living}:
	Let $\mathcal{C}\subseteq\mathbb{R}^n$ be a convex closed cone. Statistical dimension of $\mathcal{C}$ is defined as:
	\begin{align}\label{eq.statisticaldimension}
	\delta(\mathcal{C}):=\mathds{E}\|\mathcal{P}_\mathcal{C}(\bm{g})\|_2^2=\mathds{E}\mathrm{dist}^2(\bm{g},\mathcal{C}^\circ),
	\end{align}
	where $\bm{g}$ has i.i.d. standard normal distribution, and $\mathcal{P}_\mathcal{C}(\bm{x})$ is the orthogonal projection of $\bm{x}\in \mathbb{R}^n$ onto the set $\mathcal{C}$ defined as: $\mathcal{P}_\mathcal{C}(\bm{x})=\underset{\bm{z} \in \mathcal{C}}{\arg\min}\|\bm{z}-\bm{x}\|_2$.
\end{defn}
Statistical dimension specifies the boundary of success and failure in random convex programs with affine constraints.
\section{Main results}\label{section.mainresult}
In this section, we first present an upper-bound for the required number of Gaussian measurements for the case of a redundant analysis operator $\bm{\Omega}$. A lower-bound is also derived for non-singular $\bm{\Omega}$. The lower-bound is not new and was previously reported in \cite[Theorem A]{amelunxen2017effective}, but here we present a simpler approach for the proof. 
\begin{prop}\label{prop.upperboundstatis}
Let $\bm{x}\in\mathbb{R}^n$ be a $s$-analysis sparse vector with redundant analysis operator $\bm{\Omega}\in\mathbb{R}^{p\times n}$($p\ge n$). Then,
\begin{align}\label{eq.mainequation}
\delta(\mathcal{D}(\|\bm{\Omega} \cdot\|_{1,\bm{w}},\bm{x}))\le \kappa^2(\bm{\Omega})\delta(\mathcal{D}(\|\cdot\|_{1,\bm{w}},\bm{\Omega x})).
\end{align} 
Moreover, if $\bm{\Omega}$ is non-singular and $p=n$,
\begin{align}
\frac{1}{\kappa^2(\bm{\Omega})}\delta(\mathcal{D}(\|\cdot\|_{1,\bm{w}},\bm{\Omega x}))\le\delta(\mathcal{D}(\|\bm{\Omega} \cdot\|_{1,\bm{w}},\bm{x}))\le&\nonumber\\
\kappa^2(\bm{\Omega})\delta(\mathcal{D}(\|\cdot\|_{1,\bm{w}},\bm{\Omega x}))&
\end{align}
\end{prop}
Proof. See Appendix \ref{proof.l1analysis}.\\
\begin{thm}\label{thm.stablerecovery}
Let $\bm{x}\in\mathbb{R}^n$. Let the entries of $\bm{A}\in\mathbb{R}^{m\times n}$ be a random matrix with entries drawn from an i.i.d. standard normal distribution. If $\bm{y}=\bm{A x}\in\mathbb{R}^m$, and
\begin{align}\label{eq.measure_exact}
m>\left(\kappa(\bm{\Omega})\sqrt{\delta(\mathcal{D}(\|\cdot\|_{1,\bm{w}},\bm{\Omega}\bm{x})}+t)\right)^2+1,
\end{align}
for a given $t>0$, then, $\mathsf{P}_{0}^{\bm{w}}$ recovers $\bm{x}$ with probability at least $1-e^{-\frac{t^2}{2}}$. Also, if $\bm{y}=\bm{A x}+\epsilon$ and $\bm{\Omega} \bm{x}_{ap}$ is the best $\tilde{s}$-term approximation of the $s$- sparse vector $\bm{\Omega x}$ ($s\ge \tilde{s}$), then any solution $\widehat{\bm{x}}$ of $\mathsf{P}_{\eta}^{\bm{w}}$ satisfies
\begin{align}\label{eq.error_stable}
\|\widehat{\bm{x}}-\bm{x}\|_2\le \frac{2\eta}{\left(\sqrt{m-1}-\kappa(\bm{\Omega})\sqrt{\delta(\mathcal{D}(\|\cdot\|_{1,\bm{w}},\bm{\Omega }\bm{x}_{ap}))}-t\right)_{+}},
\end{align}
with probability at least $1-e^{-\frac{t^2}{2}}$.
\end{thm} 
Proof. See Appendix \ref{proof.thm.stablerecovery}.

In the exact recovery case, we determine the suitable weights by minimizing the right-hand side of (\ref{eq.measure_exact}). In the noisy setting, for stable and robust recovery, we determine the weights by minimizing the reconstruction error (the right-hand side of \ref{eq.error_stable}):
\begin{align}\label{eq.omegastar}
\bm{\omega}^*=\mathop{\arg\min}_{\bm{\nu}\in\mathbb{R}_{+}^L}\mathds{E}\mathrm{dist}^2(\bm{g},(\bm{D\nu})\odot \partial \|\cdot\|_1(\bm{\Omega x})),
\end{align}
where $\bm{D}:=[\bm{1}_{\mathcal{P}_1},..., \bm{1}_{\mathcal{P}_L}]\in\mathbb{R}^{p\times L}$. The latter optimization problem is very similar to the one in weighted $\ell_1$ minimization. With the same approach as in \cite{flinth2016optimal}, one can show that \eqref{eq.omegastar} reduces to solving the following equations simultaneously \cite[Corollary 11]{flinth2016optimal}:
\begin{align}\label{eq.suitableweights}
\alpha_i\omega^*_i=(1-\alpha_i)\sqrt{\frac{2}{\pi}}\int_{\omega^*_i}^{\infty}(u-\omega^*_i)e^{-\frac{u^2}{2}}du~:~i=1,..., L.
\end{align}
It is not obvious whether the inequality (\ref{eq.mainequation}) in Proposition \ref{prop.upperboundstatis} is tight for highly redundant and coherent analysis operators. However, numerical evidence suggests that the obtained bound is close to the $\ell_1$ analysis phase transition curve for low-redundancy regime (See Figure \ref{fig.phasebound}). 

In practice, one may encounter some inaccuracies in determining $\bm{\alpha}\in\mathbb{R}^{L}$. The study of the sensitivity of weights to the inaccuracies in $\alpha$ were previously considered in \cite{daei2018exploiting}. Fortunately, small changes in $\alpha$ are shown to have insignificant impact on the derived weights.
\section{Simulation Results}\label{section.simulation}
In this section, we numerically study the effect of weights obtained by (\ref{eq.suitableweights}) on the number of required measurements. First, we consider the scaling of the required number of measurements for successful recovery of (\ref{problem.l1analysis}) with analysis sparsity. The heatmap in Figure \ref{fig.phasebound} shows the empirical probability of success. Indeed, the results are consistent with (\ref{eq.mainequation}). In the second experiment, we generate a $s=10$-analysis sparse random vector $\bm{x}\in\mathbb{R}^{55}$ in two different analysis operators with $\kappa(\bm{\Omega})=1.1$ and $\kappa(\bm{\Omega})=230$. We consider two random sets $\mathcal{P}_1$ and $\mathcal{P}_2$ that partition the analysis domain $\{1,... ,p\}$ with $\alpha_1=\frac{7}{10}$ and $\alpha_2=\frac{3}{50}$. The suitable weights are obtained via equation (\ref{eq.suitableweights}) by MATLAB function \textsf{fzero}. Figures \ref{fig.success1} and \ref{fig.success2} show the success rate of $\mathsf{P}_{0}^{\bm{w}^*}$ averaged over $50$ Monte Carlo simulations. It is evident that the weighted $\ell_1$ analysis with suitable weights needs less number of measurements than regular $\ell_1$ analysis.

\begin{figure}[t!]
	\hspace*{-.5cm}
	\includegraphics[scale=.3]{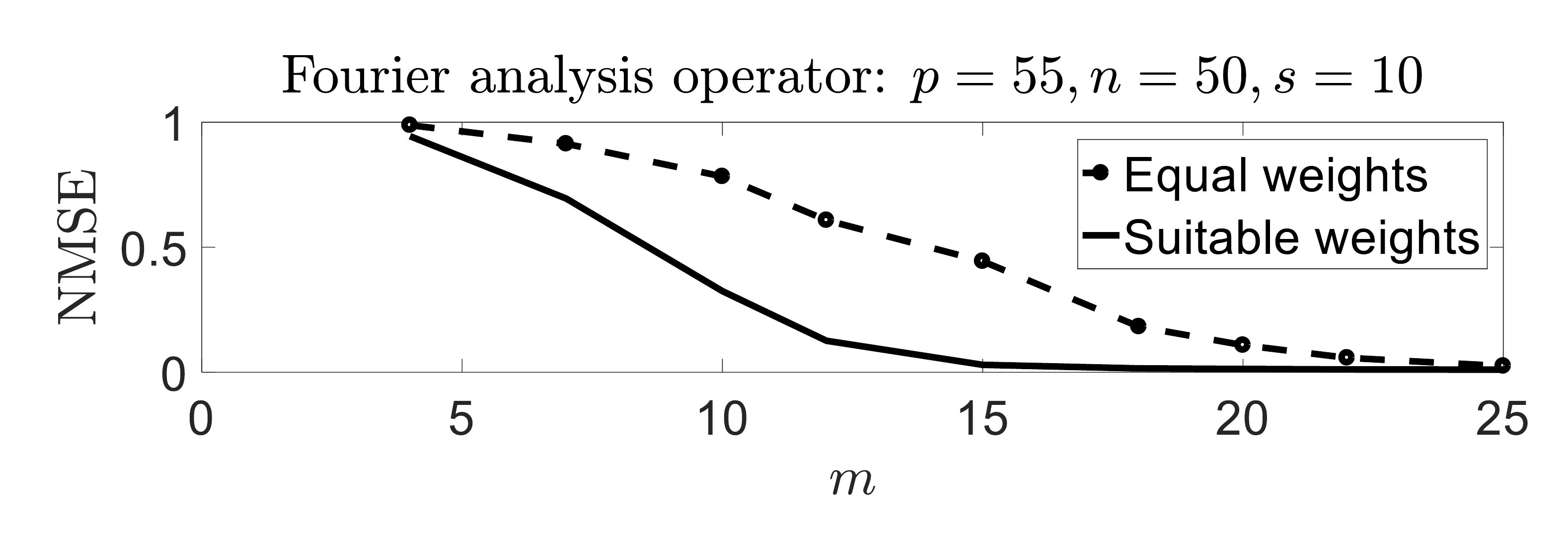}
	\caption{NMSE versus number of required measurements in problems (\ref{problem.l1analysis}) and (\ref{problem.weightedl1analysis}) in the case that the ground truth signal $\bm{x}\in\mathbb{R}^{50}$ is sparse after applying Fourier analysis operator $\bm{\Omega}\in\mathbb{C}^{55\times 50}$. NMSE is computed out of $50$ Monte Carlo simulations.} 
	\label{fig.error1}
\end{figure}

{\color{\change} In a separate scenario, we repeat the second experiment for redundant Fourier analysis operator which is widely used in line spectral and direction of arrival estimation. In this experiment, the measurements are contaminated with additive noise ($\mathrm{SNR}=30~dB$). For the recovery, both $\mathsf{P}_{\eta}$ and $\mathsf{P}_{\eta}^{\bm{w}^*}$ are implemented. Also, the random analysis support estimates $\mathcal{P}_1$ and $\mathcal{P}_2$ are such made that $\alpha_1=\frac{8}{10}$ and $\alpha_2=\frac{2}{45}$. The normalized mean square error
\begin{align}
\mathrm{NMSE}=\frac{\|\hat{\bm{x}}-\bm{x}\|_2}{\|\bm{x}\|_2},
\end{align} 
is averaged out of $50$ trials. From Figure \ref{fig.error1}, it is clear that weighted $\ell_1$ analysis with suitable weights obtained from (\ref{eq.suitableweights}) needs less measurements than regular $\ell_1$ analysis in a fixed NMSE. 

In the last experiment, we investigate a more practical scenario where a Shepp-Logan image ( denoted by $\bm{X}$ of size $n_1\times n_2$ pixels) is under-sampled with a fat random Gaussian matrix (of size $m\times n_1 n_2$) and passed through an additive noise with SNR=$10~dB$. Figure \ref{fig.mri} illustrates the recovery of this image ($n_1=n_2=128$) by solving \ref{problem.l1analysis} and \ref{problem.weightedl1analysis} when $\bm{\Omega}\in\mathbb{R}^{114688\times 16384}$ is a redundant wavelet matrix from daubechies family (the weights in (\ref{problem.weightedl1analysis}) are obtained via \ref{eq.suitableweights}) and in case of $m=6554$. The recovery problems (\ref{problem.l1analysis}) and (\ref{problem.weightedl1analysis}) are carried out using TFOCS algorithm \cite{becker2011templates}. The quality of each method is reported in terms of the Peak SNR (PSNR) given by:
\begin{align}
{\rm PSNR}(\bm{X},\widehat{\bm{X}}):=20\log_{10}\left(\frac{\|\bm{X}\|_{\infty}\sqrt{n_1n_2}}{\|\bm{X}-\widehat{\bm{X}}\|_F}\right).
\end{align}
We assume $11$ disjoint support estimators in the analysis domain with known level of contributing ($\{\alpha_i\}_{i=1}^{11}$ in \eqref{eq.parameters}) with top $10\%$ (specifying $\tilde{s}$ in Theorem \ref{thm.stablerecovery}) of wavelet coefficients. As shown by Figure \ref{fig.mri}, while $\mathsf{P}_{\eta}^{\bm{w}^*}$ has an acceptable performance with PSNR=$18.53~dB$, $\mathsf{P}_{\eta}$ clearly fails with a poor performance PSNR=$10.2~dB$.
\begin{figure}[t]
	\hspace*{-.7cm}
	\includegraphics[scale=.28]{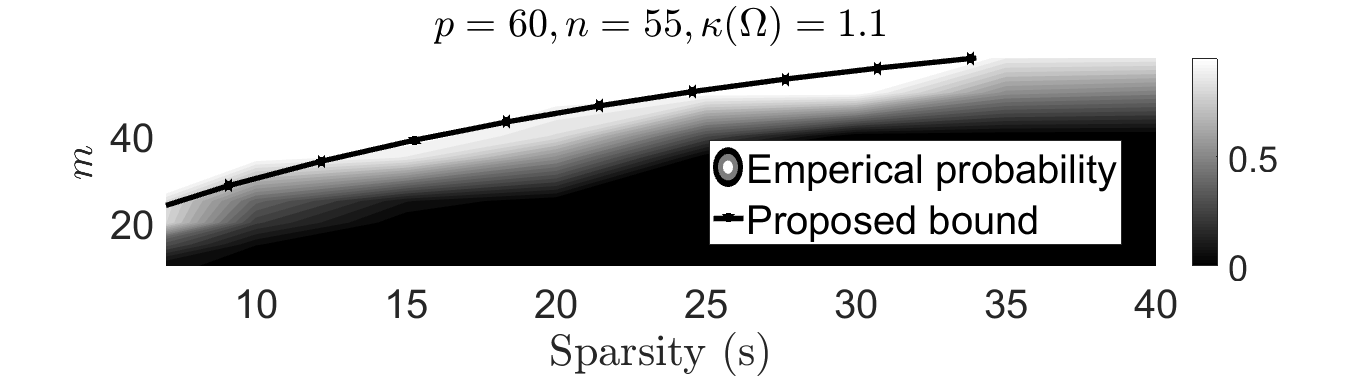}
	\caption{Empirical probability that problem (\ref{problem.l1analysis}) recovers $\bm{x}\in\mathbb{R}^{55}$ that has $s$ non-zero entries after applying a redundant analysis operator with $\kappa(\bm{\Omega})=1.1$. The black line shows the number of measurements obtained by (\ref{eq.mainequation}).} 
	\label{fig.phasebound}
\end{figure}
\begin{figure}[t]
	\hspace*{-.5cm}
	\includegraphics[scale=.3]{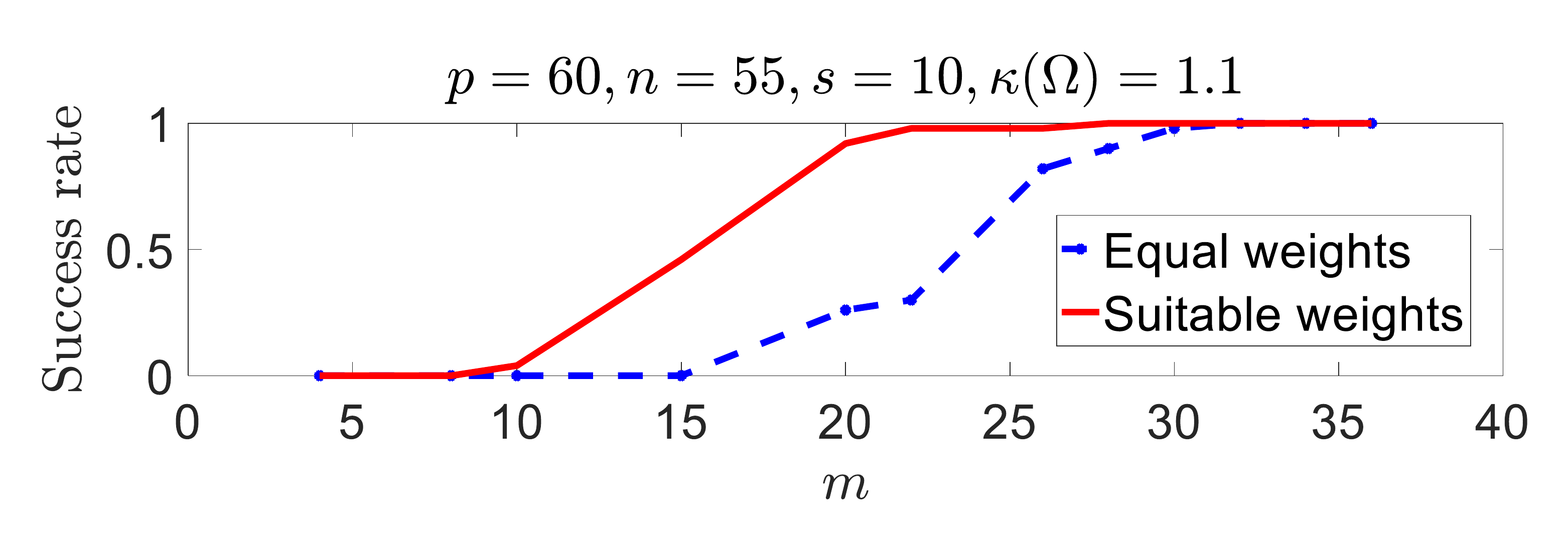}
	\caption{The probability that (\ref{problem.l1analysis}) and (\ref{problem.weightedl1analysis}) succeed to recover $\bm{x}\in\mathbb{R}^{55}$ from Gaussian linear measurements. $p=60$, $n=55$, $s=10$, $\kappa(\Omega)=1.1$. Suitable weights used in (\ref{problem.weightedl1analysis}), are obtained from (\ref{eq.suitableweights}).}
	\label{fig.success1}
\end{figure}
\begin{figure}[t]
	\hspace*{-.5cm}
	\includegraphics[scale=.3]{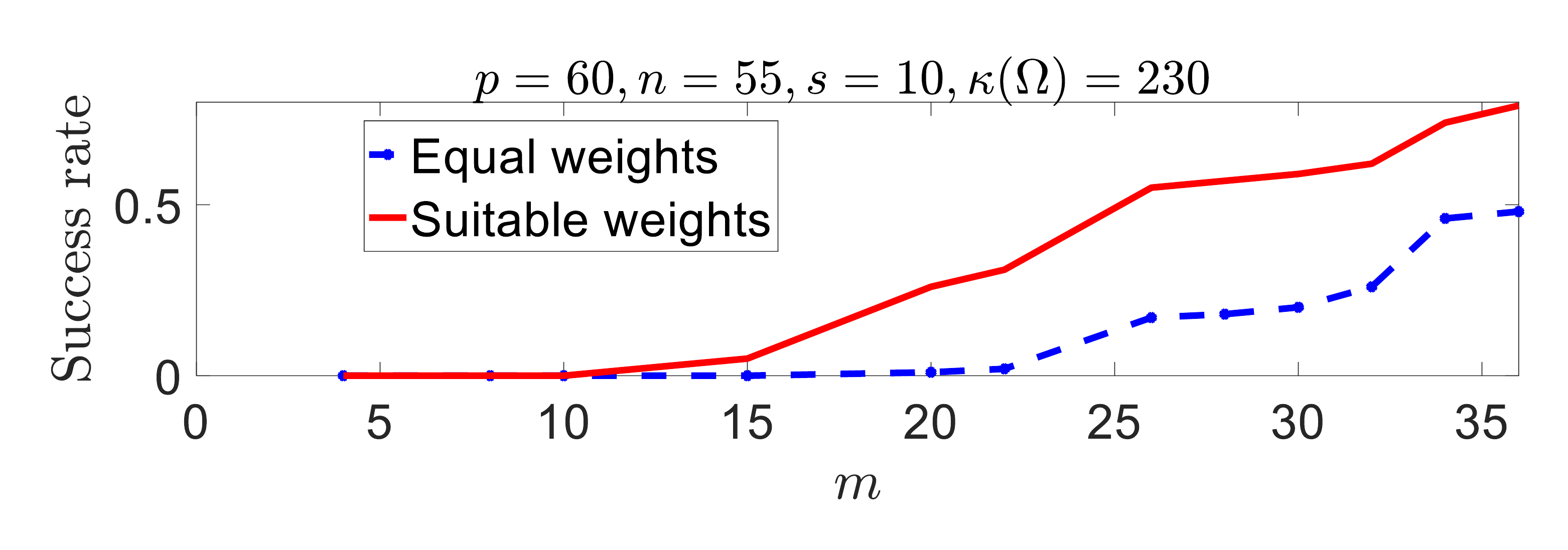}
	\caption{The probability that (\ref{problem.l1analysis})  and (\ref{problem.weightedl1analysis}) succeed to recover $\bm{x}\in\mathbb{R}^{55}$ from Gaussian linear measurements. $p=60$, $n=55$, $s=10$, $\kappa(\Omega)=230$. Suitable weights used in (\ref{problem.weightedl1analysis}), are obtained from (\ref{eq.suitableweights}).} 
	\label{fig.success2}
\end{figure}
\begin{figure*}[t]
	\centering
	\mbox{\subfigure[]{\includegraphics[width=2.34in]{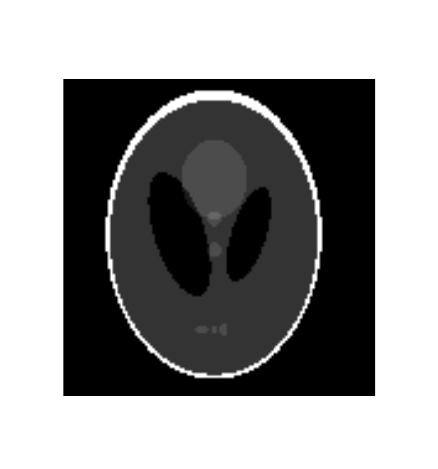}\label{fig.mri_main}}\quad
		\subfigure[]{\includegraphics[width=2.34in]{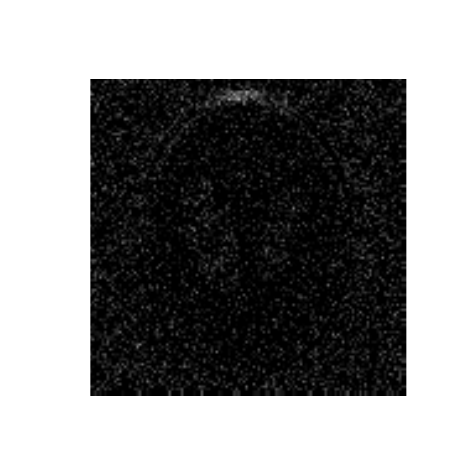}\label{fig.mri_l1ana}}\quad
		\subfigure[]{\includegraphics[width=2.34in]{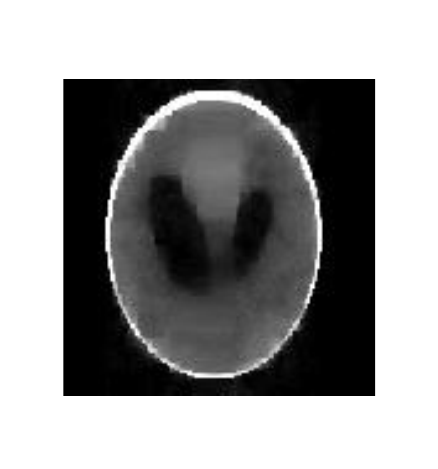}\label{fig.mri_l1anaweighted}}}
	\caption{ Effect of suitable weights in analysis sparse recovery with prior information. The associated parameters are $n_1=n_2=128$, and $m=6554$. \subref{fig.mri_main} Ground truth image (PSNR=$\infty ~dB$). \protect \subref{fig.mri_l1ana}Recovered image via $\ell_1$ analysis (PSNR=$10.2 ~dB$)  \protect \subref{fig.mri_l1anaweighted} Recovered image with weighted $\ell_1$ analysis with suitable weights (PSNR=$18.53~dB$).} \label{fig.mri}
\end{figure*}}
\appendix
\section{Proofs of Lemmas and Propositions}
\subsection{Proof of Proposition \ref{prop.upperboundstatis}}\label{proof.l1analysis}
\begin{proof}	
In the following, we relate $\mathcal{D}(\|\bm{\Omega}\cdot\|_1,\bm{x})$ to $\mathcal{D}(\|\cdot\|_1,\bm{\Omega x})$.
	\begin{align}\label{eq.descent cone relation}
	&\mathcal{D}(\|\bm{\Omega}\cdot\|_1,\bm{x})^{\circ\circ}=\mathrm{closure}(\mathcal{D}(\|\bm{\Omega}\cdot\|_1,\bm{x}))\nonumber\\
	&\mathcal{D}(\|\bm{\Omega}\cdot\|_1,\bm{x})=\mathrm{cone}^{\circ}(\bm{\Omega}^T\partial\|\cdot\|_1(\bm{\Omega x}))\nonumber\\
	&\{\bm{w}\in\mathbb{R}^n:\langle \bm{w},\bm{\Omega}^T\bm{v}\rangle\le0~:~\forall \bm{v}\in \mathrm{cone}(\partial\|\cdot\|_1(\bm{\Omega x}))\}=\nonumber\\
	&\{\bm{w}\in\mathbb{R}^n:~\bm{\Omega w}\in \mathrm{cone}^{\circ}(\partial\|\cdot\|_1(\bm{\Omega x}))\}=\nonumber\\
	&\{\bm{w}\in\mathbb{R}^n:~\bm{\Omega w}\in\mathcal{D}(\|\cdot\|_1,\bm{\Omega x})\}.
	\end{align}
	Therefore,
	\begin{align}\label{rabetesubset}
	\bm{\Omega} \mathcal{D}(\|\bm{\Omega}\cdot\|_1,\bm{x})\subset \mathcal{D}(\|\cdot\|_1,\bm{\Omega} \bm{x}).
	\end{align}
	In particular, if $\bm{\Omega}$ is non-singular and $p=n$,
	\begin{align}\label{rabeteequality}
	\bm{\Omega} \mathcal{D}(\|\bm{\Omega}\cdot\|_1,\bm{x})= \mathcal{D}(\|\cdot\|_1,\bm{\Omega} \bm{x}),
	\end{align}
	where in the last line of (\ref{eq.descent cone relation}), we used the fact that $\mathcal{D}(\|\bm{\Omega}\cdot\|_1,\bm{x})$ is a closed convex set. In the following, we state Sudakov-Fernique inequality which helps to control the supremum of a random process by that of a simpler random process and is used to find an upper-bound for $\delta(\mathcal{D}(\|\bm{\Omega}\cdot\|_1,\bm{x}))$.
	\begin{thm}\label{thm.sodakov}(Sudakov-Fernique inequality).
		Let $T$ be a set and $\mathbf{X}=(X_t)_{t\in T}$ and $\mathbf{Y}=(Y_t)_{t\in T}$ be Gaussian processes satisfying $\mathds{E}[X_t]=\mathds{E}[Y_t]~:~\forall t\in T$ and $\mathds{E}|X_t-X_s|^2\le\mathds{E}|Y_t-Y_s|^2~:~\forall s,t\in T$, then
		\begin{align}
		\mathds{E}\sup_{t\in T}X_t^2\le\mathds{E}\sup_{t\in T}Y_t^2.
		\end{align}
	\end{thm}
\vspace{-.5cm}
	\begin{align}\label{eq.Gausianwidupper}
	&\delta(\mathcal{D}(\|\bm{\Omega}\cdot\|_1,\bm{x})):=\mathds{E}\bigg(\sup_{\substack{\bm{w}\in\mathcal{D}(\|\bm{\Omega}\cdot\|_1,\bm{x})\\
			\|\bm{w}\|_2\le 1}}\langle \bm{g},\bm{w}\rangle\bigg)^2\le \|\bm{\Omega}\|_{2\rightarrow 2}^2
	\nonumber\\
	&\mathds{E}\bigg(\sup_{\substack{\bm{w}\in\mathcal{D}(\|\bm{\Omega}\cdot\|_1,x)\\
			\|\bm{\Omega}\|_{2\rightarrow 2}\|\bm{v}\|_2\le 1}}
	\langle \bm{g},\bm{v}\rangle\bigg)^2\le\|\bm{\Omega}\|_{2\rightarrow 2}^2\|\bm{\Omega}^{\dagger}\|_{2\rightarrow 2}^2\nonumber\\
	&\mathds{E}\bigg(\sup_{\substack{\bm{w}\in\mathcal{D}(\|\bm{\Omega}\cdot\|_1,x)\\
			\|\bm{\Omega}\|_{2\rightarrow 2}\|\bm{v}\|_2\le 1}}
	\langle \bm{h},\bm{\Omega}\bm{v}\rangle\bigg)^2\kappa^2(\bm{\Omega})\mathds{E}\bigg(\sup_{\substack{\bm{w}\in\mathcal{D}(\|\bm{\Omega}\cdot\|_1,\bm{x})\\
			\|\bm{\Omega}\bm{v}\|_2\le 1}}
	\langle \bm{h},\bm{\Omega v}\rangle\bigg)^2\nonumber\\
	&\le\kappa^2(\bm{\Omega})\mathds{E}\bigg(\sup_{\substack{\bm{z}\in\bm{\Omega}\mathcal{D}(\|\bm{\Omega}\cdot\|_1,\bm{x})\\
			\|\bm{z}\|_2\le 1}}
	\langle \bm{h},\bm{z}\rangle\bigg)^2\le \nonumber\\
	&\kappa^2(\bm{\Omega})\mathds{E}\bigg(\sup_{\substack{\bm{z}\in\mathcal{D}(\|\cdot\|_1,\bm{\Omega x})\\
			\|\bm{z}\|_2\le 1}}
	\langle \bm{h},\bm{z}\rangle\bigg)^2=\kappa^2(\bm{\Omega})\delta(\mathcal{D}(\|\cdot\|_1,\bm{\Omega x})),
	\end{align}
	where in (\ref{eq.Gausianwidupper}), $\bm{h}\in\mathbb{R}^p$ is a standard normal vector with i.i.d components. In the first inequality of (\ref{eq.Gausianwidupper}), we used the change of variable $\bm{v}=\|\bm{\Omega}\|_{2\rightarrow2}^{-1}\bm{w}$. The second inequality comes from Theorem \ref{thm.sodakov} with $X_{\bm{v}}=\langle \bm{g},\bm{v} \rangle$ and $Y_{\bm{v}}=\|\bm{\Omega}^\dagger\|_{2\rightarrow2}\langle \bm{h},\bm{\Omega}\bm{v} \rangle$ and the fact that:
	\begin{align}
	&\mathds{E}|X_{\bm{v}}-X_{\bm{w}}|^2=\|\bm{v}-\bm{w}\|_2^2\le \|\bm{\Omega}^\dagger\|_{2\rightarrow 2}^2\|\bm{\Omega}(\bm{v}-\bm{w})\|_2^2 =\nonumber\\
	&=\mathds{E}|Y_{\bm{v}}-Y_{\bm{w}}|^2~:~\forall \bm{v},\bm{w}\in\mathbb{R}^n.
	\end{align} 
	The last inequality comes from (\ref{rabetesubset}). In the special case $p=n$ and $\bm{\Omega}$ is non-singular we have:
	\begin{align}
	&\delta(\mathcal{D}(\|\bm{\Omega}\cdot\|_1,\bm{x})):=
	\mathds{E}\bigg(\sup_{\substack{\bm{w}\in\mathcal{D}(\|\bm{\Omega}\cdot\|_1,\bm{x})\\
			\|\bm{w}\|_2\le 1}}\langle \bm{g},\bm{w}\rangle\bigg)^2= \nonumber\\	
	&\mathds{E}\bigg(\sup_{\substack{\bm{v}\in\mathcal{D}(\|\cdot\|_1,\bm{\Omega} \bm{x})\\
			\|\bm{\Omega}^\dagger \bm{v}\|_2\le 1}}\langle \bm{g},\bm{\Omega}^\dagger\bm{v}\rangle\bigg)^2\ge\nonumber\\
	& \|\bm{\Omega}\|_{2\rightarrow 2}^{-2}\mathds{E}\bigg(\sup_{\substack{\bm{v}\in\mathcal{D}(\|\cdot\|_1,\bm{\Omega} \bm{x})\\
			\|\bm{\Omega}^\dagger \bm{v}\|_2\le 1}}\langle \bm{h},\bm{v}\rangle\bigg)^2\ge\nonumber\\
	&\|\bm{\Omega}\|_{2\rightarrow 2}^{-2}\mathds{E}\bigg(\sup_{\substack{\bm{v}\in\mathcal{D}(\|\cdot\|_1,\bm{\Omega} \bm{x})\\
			\|\bm{v}\|_2\le \|\bm{\Omega}^\dagger\|_{2\rightarrow 2}^{-1}}}\langle \bm{h},\bm{v}\rangle\bigg)^2=\frac{1}{\kappa^2(\bm{\Omega})}\delta(\mathcal{D}(\|\cdot\|_1,\bm{\Omega} \bm{x})),
	\end{align} 
	where the first inequality comes from $\bm{\Omega}^\dagger\bm{\Omega}=\bm{I}$ and (\ref{rabeteequality}). The second inequality comes from Theorem \ref{thm.sodakov} with $X_v=\langle \bm{g}, \bm{\Omega}^\dagger \bm{v}\rangle$ and $Y_v=\|\bm{\Omega}\|_{2\rightarrow 2}^{-1}\langle \bm{h},\bm{v}\rangle$ and the fact that
	\begin{align}
	&\mathds{E}|X_{\bm{v}}-X_{\bm{w}}|^2=\|\bm{\Omega}^\dagger(\bm{v}-\bm{w})\|_2^2\ge \|\bm{\Omega}\|_{2\rightarrow 2}^{-2}\|\bm{v}-\bm{w}\|_2^2 =\nonumber\\
	&=\mathds{E}|Y_{\bm{v}}-Y_{\bm{w}}|^2~:~\forall \bm{v},\bm{w}\in\mathbb{R}^n,
	\end{align}
	where the last inequality is a result of norm properties.   
\end{proof}
	\subsection{Proof of theorem \ref{thm.stablerecovery} }\label{proof.thm.stablerecovery}
	Let $T_0$ be the index set of $\tilde{s}$ largest analysis coefficients. Then, it holds that,
	\begin{align}
	\|\bm{\Omega}\bm{x}_{ap}\|_1:=\|(\bm{\Omega x})_{T_0}\|_1\ge \frac{\tilde{s}\|\bm{\Omega x} \|_1}{s},
	\end{align}
	and as a result, we have, $\mathcal{D}(\|\bm{\Omega}\cdot\|_1,\bm{x})\subseteq \mathcal{D}(\|\bm{\Omega}\cdot\|_1,\frac{s}{\tilde{s}}\bm{x})$ and thus $\delta(\mathcal{D}(\|\bm{\Omega}\cdot\|_1,\bm{x}))\le \delta(\mathcal{D}(\|\bm{\Omega}\cdot\|_1,\frac{s}{\tilde{s}}\bm{x}))$. The result in theorem \ref{thm.stablerecovery} follows from \cite[Corollary 3.5]{tropp2015convex}, Proposition \ref{prop.upperboundstatis}, and the fact that $\delta(\mathcal{D}(\|\cdot\|_1,\bm{\Omega}\bm{x}))$ only depends on the support of $\bm{\Omega}\bm{x}$.
\ifCLASSOPTIONcaptionsoff
  \newpage
\fi

\bibliographystyle{ieeetr}
\bibliography{mypaperbibe}
\end{document}